\documentclass[aps,prl,preprint,superscriptaddress,amsmath,amssymb]{revtex4}
\usepackage{graphicx}
\usepackage{color}
\begin{document}

\title{Intense 2-cycle laser pulses induce time-dependent bond-hardening in a polyatomic molecule}

\author{K. Dota}
\affiliation{Tata Institute of Fundamental Research, 1 Homi Bhabha Road, Mumbai 400 005, India}
\author{M. Garg}
\affiliation{Indian Institute of Science Education and Research Kolkata, Mohanpur 741 252, India}
\author{A. K. Tiwari}
\affiliation{Indian Institute of Science Education and Research Kolkata, Mohanpur 741 252, India}
\author{J. A. Dharmadhikari}
\affiliation{Tata Institute of Fundamental Research, 1 Homi Bhabha Road, Mumbai 400 005, India}
\author{A. K. Dharmadhikari}
\affiliation{Tata Institute of Fundamental Research, 1 Homi Bhabha Road, Mumbai 400 005, India}
\author{D. Mathur}
\affiliation{Tata Institute of Fundamental Research, 1 Homi Bhabha Road, Mumbai 400 005, India}

\begin{abstract}
A time-dependent bond-hardening process is discovered in a polyatomic molecule (tetramethyl silane, TMS) using few-cycle pulses of intense 800 nm light. In conventional mass spectrometry, symmetrical molecules like TMS do not exhibit a prominent molecular ion (TMS$^+$) as unimolecular dissociation into [Si(CH$_3$)$_3]^+$ proceeds very fast. Under strong field and few-cycle conditions, this dissociation channel is defeated by time-dependent bond-hardening: a field-induced potential well is created in the TMS$^+$ potential energy curve that effectively traps a wavepacket. The time-dependence of this bond hardening process is verified using longer-duration ($\geq$ 100 fs) pulses; the relatively ``slower" fall-off of optical field in such pulses allows the initially trapped wavepacket to leak out, thereby rendering TMS$^+$ unstable once again. Our results are significant as they demonstrate (i) optical generation of polyatomic ions that are normally inaccessible and (ii) optical control of dynamics in strong fields, with distinct advantages over weak-field control scenarios that demand a narrow bandwidth appropriate for a specified transition.
\end{abstract}

\maketitle

Using lasers to image and control complex chemical reactions remains a Holy Grail of contemporary molecular physics \cite{intro}. Availability of few-cycle laser pulses and, more recently, of attosecond pulses, is beginning to allow time-dependent nuclear and electron dynamics to be probed, enhancing our ability to gain proper insights into laser control of quantum systems. Pump-probe experiments with few-cycle pulses have succeeded in real-time mapping of nuclear wavepackets in molecules \cite{nuclear}. However, all hitherto reported work has focused on small molecules like H$_2$ and H$_2^+$.  We report here results of an experiment in which a polyatomic molecule is exposed to intense  laser pulses such  that, in the ultrashort  domain accessed when 5 fs (2-cycle) and 22 fs (8 cycles) pulses are used, a bond hardening process is directly observed while, when longer pulses ($\geq$100 fs) of approximately the same intensity are used, the bond hardening disappears. The  molecule of interest in our studies is tetramethyl  silane (TMS), a tetrahedral molecule with a Si-atom surrounded by four methyl groups [Si(CH$_3$)$_4$]. TMS is used as aviation fuel and as an internal standard for NMR instrumentation; it has many potential uses in the incipient photonics industry. As in the case of other highly symmetrical polyatomic molecules,  like SF$_6$, the molecular ion of TMS does not show up prominently in mass spectra \cite{TMS}:  Jahn-Teller distortion ensures that TMS$^+$ is unstable. The most prominent peak in mass spectra is the  [Si(CH$_3$)$_3]^+$ peak. We show in our experiments that the use of an intense few-cycle pulse helps defeat Jahn-Teller effects and TMS$^+$ can be produced as a consequence of CEP-independent bond hardening. This bond hardening is itself defeated when longer pulses (100 fs and longer) of about the same intensity are used. We  utilize time-dependent wavepacket dynamics on potential energy surfaces of the TMS-cation to rationalize our observations of such time-dependent bond hardening which, to our knowledge, has not been previously reported. Work on bond hardening, both experimental and theoretical, has focused exclusively on diatomics \cite{small} (for a recent review, see \cite{williams}). Our observations are of significance from the viewpoint of optically generating polyatomic molecular ions that are normally not accessible. Our use of intense pulses at kHz repetition rate permits us to demonstrate some measure of optical control of molecular dynamics in the strong-field regime. This offers the distinct advantage over conventional weak-field control scenarios in that a narrow bandwidth appropriate for a single, specified  transition is no longer mandatory in our strong-field scheme,  offering possibilities of dynamically modifying molecular potential surfaces involved in a wide range of chemical reactions. Our results may also stimulate theoretical interest. Bond hardening has hitherto been probed using methods based on the periodicity of the optical field; such methods are difficult to apply to 2-cycle pulses, making it necessary to develop different approaches, like applying Floquet methods to the pulse envelope rather than the carrier [C.D. Lin - private communication]. 

Our experiments on strong-field ionization of TMS were conducted using two Ti:Sapphire  femtosecond laser systems. In our CEP-stabilized system, output pulses from a Ti:S oscillator were amplified in a 4-pass amplifier at 75 MHz repetition rate, stretched to $\sim$200 ps and passed through a programmable acousto-optic dispersive filter that permits control of pulse shape and duration.  The output passed via an electro-optical modulator (which down-converted to 1 kHz repetition rate) to a 5-pass amplifier and compressor. The resulting 22 fs pulse was further compressed to 5 fs using a 1m-long Ne-filled hollow fiber and chirped dielectric mirrors. CEP stabilization was accomplished via a fast-loop in the oscillator and a slow-loop in the amplifier \cite{krause}. Our experiments with 100-200 fs pulses utilized a second Ti:S system which has been described elsewhere \cite{apb2005}. Linearly-polarized laser pulses from either of these systems were transmitted to our molecular beam apparatus \cite{earlier} in which ionization was monitored using linear time-of-flight techniques. Data acquisition at 1 kHz was in list mode using a 2.5 GHz oscilloscope operating in sequence mode. 

\begin{figure}
\includegraphics[width=7cm]{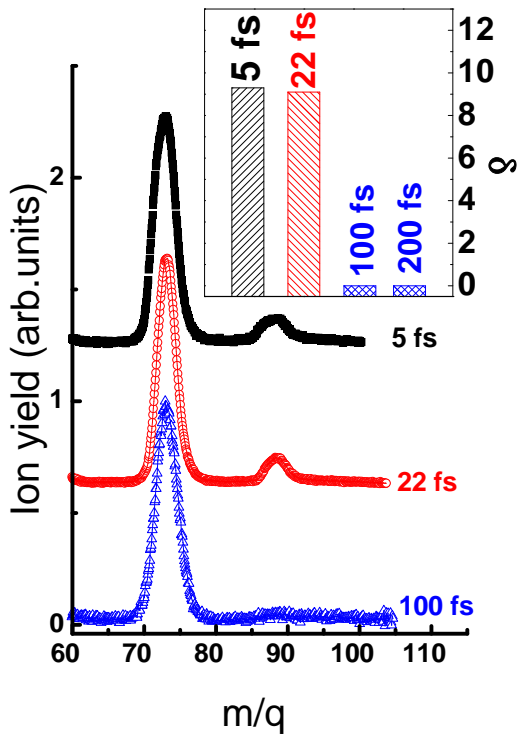}
	\caption{(Color online) Time-of-flight spectra showing [Si(CH$_3$)$_3]^+$ (m/q = 73) and TMS$^+$ (m/q = 88) obtained using three different pulse durations of about the same peak intensity ($\sim$2$\times$10$^{15}$ W cm$^{-2}$). Note the absence of the parent ion peak (m/q = 88) in the case of the 100 fs pulse. The inset shows the percentage yield, $\delta$, of the TMS$^+$ ion with respect to the [Si(CH$_3$)$_3]^+$ fragment for four pulse durations.}
\end{figure}

Typical molecular ionization patterns obtained with 5 fs, 22 fs and 100 fs pulses of peak intensity $\sim$2 PW cm$^{-2}$ are shown in Fig. 1. Consistent with the results of conventional mass spectrometry, we find that measurements made with 100 fs and 200 fs pulses yield no evidence of the TMS$^+$ molecular peak (at m/z=88). The most prominent mass spectral peak corresponds to [Si(CH$_3$)$_3]^+$ (at m/q=73) in this case \cite{width}. However, TMS$^+$ makes an appearance when a 22 fs pulse is used, and persists when 5 fs pulses are used, with the percentage ratio of the yield of (TMS)$^+$ to that of [Si(CH$_3$)$_3]^+$ remaining at $\sim$9\%.

We rationalize these observations by carrying out {\it ab initio} calculations of the field-dressed potential energy (PE) curves of TMS$^+$ obtained by scanning along one particular Si-C bond of TMS$^+$ in its ground electronic state and, then, exciting the ground vibrational wave function from neutral TMS to the TMS$^+$ ground electronic state and probing its time-dependent dynamics. Our quantumchemical computations of PE curves were carried out {\it ab initio} using the MOLPRO suite of programs \cite{molpro} with Hartree-Fock method and 6-311G** basis set that included $d$- and $f$-type polarization functions. An earlier, field-free quantumchemical study \cite{TMS} showed that the lowest vertically accessible TMS$^+$ state lies above the unimolecular dissociation limit. There also exists an adiabatic TMS$^+$ state, lying 0.6 eV below the dissociation limit, which requires one Si-CH$_3$ bond to be substantially elongated. Such a state is unlikely to be accessed in experiments that demand vertical ionization. Our dressed PE curves were computed under the influence of static fields and, in the case of TMS$^+$, show a minimum (Fig. 2) which indicates bond hardening. We show that the bond-hardening dynamics is best discussed in two stages. Population of the field-dressed TMS$^+$ PE curve is the first stage of the overall dynamics: the second stage concerns the leaking of the wavepacket from the dressed PE curve. Time-dependent bond-hardening manifests itself in the second stage.

\begin{figure}
\includegraphics[width=10cm]{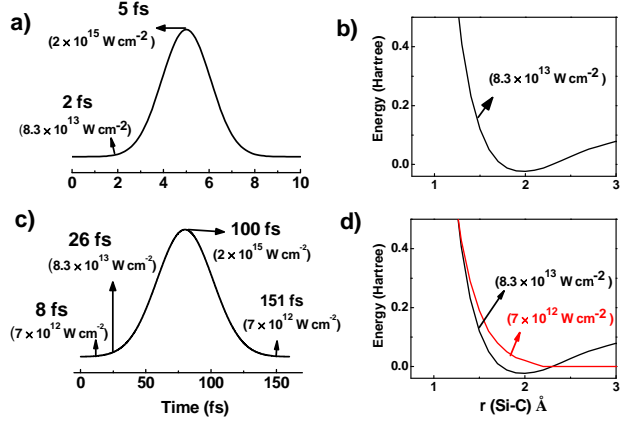}
	\caption{(Color online) a) Time evolution of a 5 fs pulse; b) An adiabatic potential well is generated along a Si-C bond after 2 fs (when the laser intensity is 83 TW cm$^{-2}$) that aids the formation of TMS$^+$ ions; c) Time evolution of a 100 fs pulse; d) Optical-field-dressed potential energy functions of TMS$^+$ (cut along the Si-C bond) at two different time positions along the rising edge of the pulse (see text).}
\end{figure}

In the first stage, we find that, for pulses of 100 fs or longer duration, the rising edge of the laser pulse ($\leq$25 fs) produces optical fields that are not strong enough to significantly dress the potential energy (PE) curve of the ground electronic state of TMS$^+$. We illustrate this with the aid of Fig. 2 which depicts the temporal evolution of a 5 fs (Fig. 2a) and a 100 fs (Fig. 2c) laser pulse whose peak intensity is 2 PW cm$^{-2}$. Corresponding sections of the dressed TMS$^+$ PE curve along the Si-C bond length are also shown (Figs. 2b,d).  In the rising edge of the 100 fs laser pulse, after 8 fs, the intensity is 7 TW cm$^{-2}$ (Fig. 2c) and the shape of the field-dressed PE curve continues to be repulsive: there is no bond-hardening and barrierless dissociation along the Si-C bond is the favored route (Fig. 2d). This lack of a potential barrier rationalizes the non-observation of TMS$^+$. It is only after 26 fs, when the intensity has increased to 83 TW cm$^{-2}$, that the optical field becomes sufficient to affect the shape of the PE curve such that bond-hardening sets in: a barrier against dissociation is formed (Fig. 2d). 

\begin{figure}
\includegraphics[width=10cm]{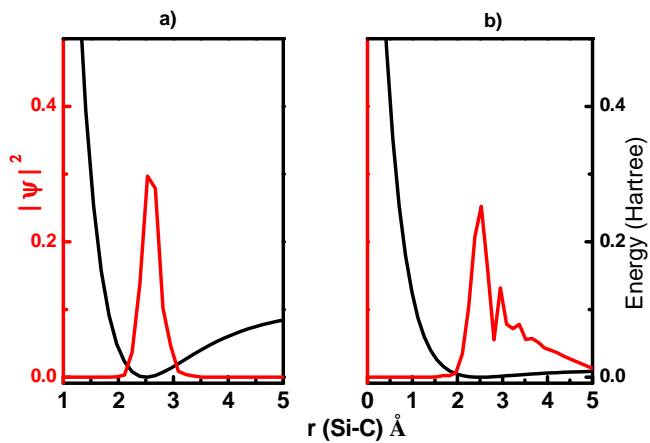}
	\caption{(Color online) a) Wavepacket in the adiabatic well of a field-dressed PE curve of TMS$^+$. The wavepacket remains trapped for a 5 fs pulse; (b) For 100 fs pulses, the wavepacket leaks out as the adiabatic well collapses in the long trailing edge of the pulse.}
\end{figure}

However, in the case of a 5 fs pulse, it requires a mere 2 fs to attain an intensity of 83 TW cm$^{-2}$ (Fig. 2a), which is sufficient to field-distort the PE curve so as to form a barrier against unimolecular dissociation (Fig. 2b): under these circumstances the Si-C bond is sufficiently hardened to allow TMS$^+$ to be formed. We find that for 22 fs pulse duration, the field-induced bond hardening occurs on sufficiently fast time scales to allow the formation of a long-lived molecular cation. What happens thereafter constitutes the second step of the time-dependent bond hardening dynamics and is depicted in schematic form in Fig. 3 which shows results of our computations of the time evolution of a wavepacket on the TMS$^+$ PE curve. The TMS$\rightarrow$TMS$^+$ excitation was obtained by Franck-Condon transition of the ground vibrational wavefunction localized in one particular Si-C bond of TMS. This wavefunction was then propagated on the field-dressed PE curve of TMS$^+$. The time evolution of wavefunction, $\psi(t)$, was carried out by numerically propagating the time-dependent Schr\"odinger equation in a series of discretized time steps, $\Delta t$,
\begin{eqnarray}
\psi(t+\Delta t) = e^{-i{\hat H}\Delta t/\hbar}\psi(t),
\end{eqnarray}
where ${\hat H}$ is the time-independent Hamiltonian. We employed the usual split-operator method  to evolve the wave function in time. Further, since the kinetic energy operator is local in momentum space, we employed the fast Fourier transform method \cite{split} to compute the action of the kinetic energy operator on the wave function.    

For bond hardening in the case of 5 fs and 22 fs pulses, we projected the lowest-energy vibrational wavepacket from the ground electronic state of neutral TMS to the ground state of the TMS$^+$ cation (Fig. 3a). The wavepacket is seen to be trapped within the potential well; this constitutes the first part of the two-stage dynamics. In the second stage, we consider the fall-off of the optical field that causes bond-hardening. The fall-off of the 5 fs laser pulse is so sharp that the wavepacket remains trapped within the potential well in the field-dressed TMS$^+$ PE curve (Fig. 3a). Indeed, the fall off of this pulse is so fast that the potential well does not collapse on this timescale. That there is no opportunity for the trapped wavepacket to leak out of the well is depicted in a movie clip \cite{movie1}. The non-adiabatic projection of the vibrational wavepacket onto the field-free PE curves that arise upon turn-off of the few-cycle optical field might even lead to enhancement of the population of bound TMS$^+$ levels. Hence, in the case of few-cycle pulses, TMS$^+$ survives long enough to enable extraction from the laser-TMS interaction zone and, thence, to be mass spectrometerically analyzed and detected.   

It would be of interest to determine the lifetime of this TMS$^+$, perhaps employing a storage ring 
as in earlier measurements of lifetimes of molecular dications \cite{ring}. In the case of longer-duration pulses ($\geq$100 fs) the first stage of the dynamics proceeds as in the 5 fs case, albeit on longer time durations. However, the second stage is now very different: the fall-off of the optical field is now adiabatic so as to permit the wavepacket to leak out \cite{movie2}, as depicted in Fig. 3b. Indeed, the slow fall-off of the pulses allows the potential well to collapse: as the pulse ends there is no well and, thus, no TMS$^+$ signal is detected in our spectrometer. Our simple model captures the essence of the dynamics that we observe; more rigorous theoretical work will require solution of a coupled electronic-nuclear Schr\"odinger equation in which the time-dependence of the optical field is explicitly accounted for. 

\begin{figure}
\includegraphics[width=7cm]{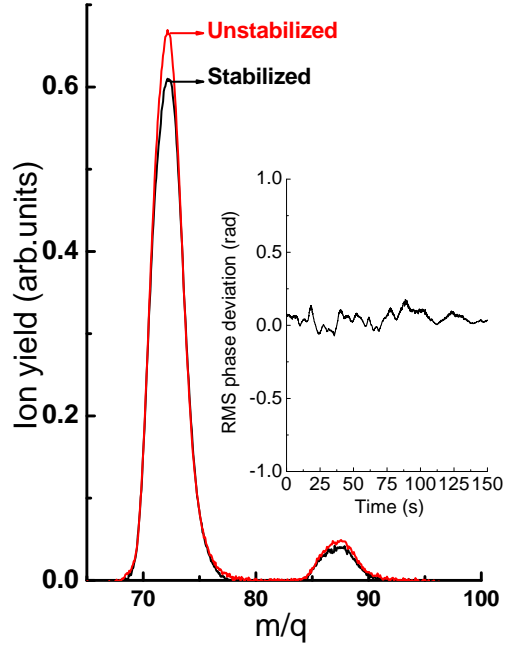}
	\caption{(Color online) TOF spectra obtained with CEP-unstabilized and CEP-stabilized 5 fs pulses. Note that the ion yield fraction TMS$^+$/[Si(CH$_3$)$_3]^+$ does not depend on whether the CEP is random or stabilized. Inset shows the rms phase stability (177 mrad) over a period 3 times longer than the data acquisition time.}
\end{figure}

In our experiments with few-cycle pulses, does the bond hardening process depend on carrier envelope phase? Our preliminary experiments to address this question indicate (Fig. 4) that the ratio of (TMS)$^+$ to [Si(CH$_3$)$_3]^+$ yields (as a fraction of total ion yield) is essentially independent of whether the CEP is random or stabilized. While the ionization channel seems to be CEP-independent, further work is required to probe CEP effects on different fragmentation channels and to, thus, explore how CEP affects the time-dependent bond hardening process in TMS.  

In summary, we have experimentally demonstrated a time-dependent bond hardening process in a polyatomic molecule (tetramethyl silane, TMS) which, under conditions prevailing in conventional mass spectrometry, does not exhibit a stable molecular ion (TMS$^+$) as unimolecular dissociation into [Si(CH$_3$)$_3]^+$ proceeds very fast. However, in strong field and few-cycle conditions, this dissociation channel is defeated due to a time-dependent bond hardening process: a field-induced potential well is created in the TMS$^+$ PE curve that is sufficient to effectively trap a wavepacket and retain it. The time-dependence of this bond hardening process is verified in experiments with longer-duration ($\geq$ 100 fs) pulses in which the relatively ``slower" fall-off of pulse intensity (optical field) allows the initially trapped wavepacket to leak out. Dynamical modification of PE surfaces due to the type of bond hardening that we demonstrate here might have wider implications, like inducing long-lived dications such as H$_2$O$^{2+}$ which are normally unstable.        
 
We acknowledge useful discussions with Hema Ramachandran and Rajesh Vatsa.

\end{document}